\documentstyle[aps]{revtex}

\begin{document}
\title{LARGE\ ORBITAL MAGNETIC MOMENT\ IN\ FeBr$_2$ }
\author{Z.Ropka}
\address{C{enter for Solid State Physics, \'{S}w. Filip 5, 31-150 Krak\'{o}w, POLAND. 
}}
\author{R. Michalski}
\address{{Center for Solid State Physics, \'{S}w. Filip 5, 31-150 Krak\'{o}w.}\\
{Inst. of Physics, Pedagogical University, 30-084 Krak\'{o}w, POLAND.}}
\author{R. J.Radwanski}
\address{{Center for Solid State Physics, \'{S}w. Filip 5, 31-150 Krak\'{o}w.}\\
{Inst. of Physics, Pedagogical University, 30-084 Krak\'{o}w, POLAND.}}
\maketitle

\begin{abstract}
Magnetic moment of the Fe$^{2+}$ ion in FeBr$_2$ has been calculated to be
4.32 $\mu _B$ in the magnetically-ordered state at 0 K. It is composed from
the spin moment of 3.52 $\mu _B$ and the orbital moment of +0.80 $\mu _B.$
These calculations show that for the meaningful analysis of e-m properties
of FeBr$_2$ the spin-orbit coupling is essentially important and that the
orbital moment is largely unquenched.

PACS 75.10.Dg 71.70.Ch 71.28+d 75.30.Mb

Keywords: crystal-field interactions, spin-orbit coupling, Fe$^{2+}$ ion,
FeBr$_2$
\end{abstract}

\pacs{75.10.Dg, 71.70.Ch, 71.28+d, 75.30.Mb}

FeBr$_2$ exhibits the strong metamagnetic properties [1,2]. At 4.2 K an
external field of 3.15 T causes a jump of the magnetization from the almost
zero value to a very big value of 110 emu/g [1,2]. This latter value
corresponds to 4.4 $\mu _B$ per the iron-ion moment. This value exceeds a
theoretical value of 4.0 $\mu _B$ expected for the spin-only moment. The
explanation of this large magnetic moment is a subject of long-time debate
within the 3d-magnetism theoretisians [3-9].

We have calculated the atomic-like magnetic moment of the Fe$^{2+}$ ion in
the ionic FeBr$_2$ compound to be 4.32 $\mu _B$ in the magnetically-ordered
state at 0 K. This moment is composed from the spin moment of 3.52 $\mu _B$
and the orbital moment of +0.80 $\mu _B$. In the paramagnetic state the
total moment amounts to 4.26 $\mu _B$ (3.48 and 0.78 $\mu _B$).

The calculations have been performed within the quantum atomistic
solid-state theory, that points out the existence of the discrete energy
spectrum in the 3d-ion containing compounds associated with quasi-atomic 3d
states. We have considered 6 d electrons of the Fe$^{2+}$ ion to form the
highly-correlated atomic like electronic system 3d$^6$. Such the system is
characterized by L=2 and S=2 [10-12]. We have taken into account the crystal
field (CEF)\ interactions (cubic CEF parameter B$_4$ = +200 K, off-cubic
trigonal distortion parameter B$_2^0$= -30 K) and the intra-atomic
spin-orbit coupling ($\lambda _{s-o}$ = -150 K).

The crystal-field interactions in the hexagonal structure of FeBr$_2$ cause
the Fe-ion moment to lie along the hexagonal axis. The hexagonal axis is the
trigonal axis of the local Br$^{1-}$ octahedron. As a consequence the
antiferromagnet FeBr$_2$ is very good example of the Ising-type magnetic
compound. The metamagnetic transition is associated with the reversal of the
AF moment. The Fe$^{2+}$-ion moment strongly depends on the local symmetry.
The local symmetry is reflected in the symmetry of the CEF. The
low-temperature magnetic moment can vary from the zero value to almost 6 $%
\mu _B$ as the function of the local surrounding of the Fe$^{2+}$ ion. The
zero value is the manifestation of the Kramers and Jahn-Teller theorems for
the highly-correlated electronic system with the even number of involved
electrons.

In conclusion, the Fe$^{2+}$ ion in FeBr$_2$ exhibits large atomic-like
magnetic moment with the very substantial orbital magnetic moment.The
derived electronic structure exhibits the discrete energy spectrum, like in
Refs 6 and 13 (the Co$^{3+}$ ion is equivalent to the Fe$^{2+}$ ion), in
contrary to the presently-in-fashion results with the continuous energy
spectrum for the 3d states [7-9].

{\bf Figure captions:}Fig.1. The calculated temperature dependence of the
local magnetic moment of the Fe$^{2+}$ ion in FeBr$_2$ together with the
spin and orbital contributions.

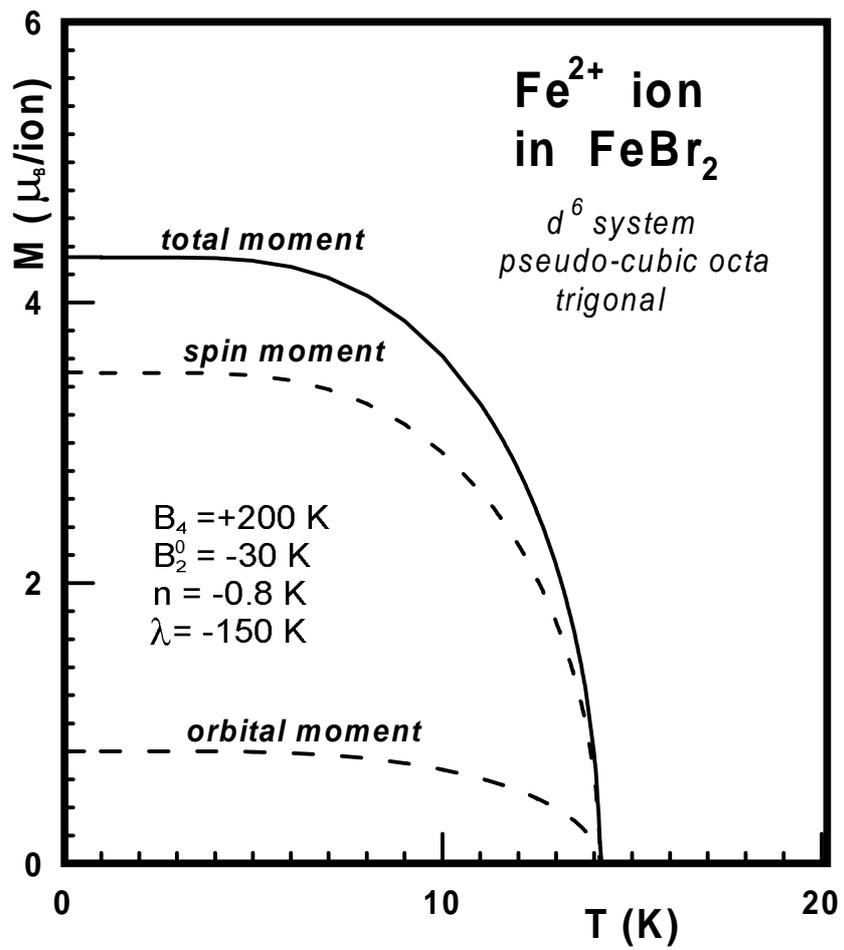
\end{document}